\documentclass[pre,twocolumn,showpacs]{revtex4}
\usepackage{graphicx,amsmath,amssymb}

\newcommand{\figwidth}{0.90\columnwidth}
\newcommand{\eq}[1]{Eq.(\ref{#1})}
\newcommand{\fig}[1]{Fig.~\ref{#1}}
\newcommand{\avg}[1]{{\langle #1 \rangle}}
\newcommand{\olcite}[1]{Ref.~\onlinecite{#1}}

\newcommand{\kb}{ k_{\rm B} }
\newcommand{\rhocp}{ \rho_{\rm cp} }
\newcommand{\riso}{ \rho_{\rm ISO} }
\newcommand{\rnem}{ \rho_{\rm NEM} }

\newcommand{\Pn}{ P(N) }
\newcommand{\gin}{ \gamma_{\rm IN} }

\newcommand{\lx}{ L_{\rm x} }
\newcommand{\ly}{ L_{\rm y} }
\newcommand{\lz}{ L_{\rm z} }

\begin{document}

\title{Interfacial tension of the isotropic--nematic interface in
  suspensions of soft spherocylinders}

\author{R. L. C. Vink and T. Schilling}

\affiliation{Institut f\"ur Physik, Johannes Gutenberg-Universit\"at,
D-55099 Mainz, Staudinger Weg 7, Germany}

\date{\today}

\begin{abstract} The isotropic to nematic transition in a system of soft
spherocylinders is studied by means of grand canonical Monte Carlo
simulations. The probability distribution of the particle density is used
to determine the coexistence density of the isotropic and the nematic
phases. The distributions are also used to compute the interfacial tension
of the isotropic--nematic interface, including an analysis of finite size
effects. Our results confirm that the Onsager limit is not recovered until
for very large elongation, exceeding at least $L/D=40$, with $L$ the
spherocylinder length and $D$ the diameter. For smaller elongation, we
find that the interfacial tension increases with increasing $L/D$, in
agreement with theoretical predictions.\end{abstract}


\pacs{61.20.Ja,64.75.+g}

\maketitle

\section{Introduction}

On change of density, suspensions of rod--like particles undergo a phase
transition between an isotropic fluid phase, where the particle
orientations are evenly distributed, and an anisotropic nematic fluid
phase, where the particle orientations are on average aligned. This
phenomenon was explained by Onsager in a theory based on infinitely
elongated hard spherocylinders \cite{onsager:1949}. Onsager theory has
been remarkably successful at describing the isotropic to nematic (IN)
transition, and still serves as the basis for many theoretical
investigations of the properties of liquid crystals. Over the last twenty
years, for instance, several groups have investigated the properties of
the IN interface using Onsager--type density functional approaches
\cite{doi:kuzuu:1985, mcmullen:1988, chen.noolandi:1992, chen:1993,
koch.harlen:1999, shundyak.roij:2001}. An important finding of these
studies is that the interfacial tension $\gin$ of the IN interface is
minimized when the director, which is the axis of average orientation of
the particles, lies in the plane of the interface. In the case of
in--plane alignment, $\gin$ is predicted to be very low, but the precise
value varies considerably between different authors \cite{schoot:1999,
chen.gray:2002}. Theoretical estimates for $\gin$ typically range from
0.156 \cite{shundyak.roij:2001} to 0.34 \cite{mcmullen:1988}, in units of
$\kb T / LD$, with $L$ the rod length, $D$ the rod diameter, $T$ the
temperature, and $\kb$ the Boltzmann constant.

Obviously, the Onsager limit of infinite rod length is purely academic. In
order to describe more realistic situations, it is necessary to go beyond
the Onsager approximation, and consider the case of finite rod length. An
example is the theoretical work of \olcite{velasco.mederos:2002}, which
demonstrates that the interfacial tension in the case of finite rod length
is considerably lower than predicted by Onsager theory.

To test the accuracy of the theoretical estimates of $\gin$, one might
envision a direct comparison to experimental data. Unfortunately, this is
not straightforward. The models used in theoretical treatments of the IN
interface are typically rather simplistic, usually based on a
short--ranged pair potential in a system of monodisperse spherocylinders.
It is not reasonable to expect quantitative agreement with experiments
using these models, because the interactions in the experimental system
will be much more complex. For example, polydispersity may be an important
factor, and it is not clear to what extent long--range interactions play a
role. Even the experimental determination of the rod dimensions $L$ and
$D$, required if a comparison to theory is to be made, presents
complications \cite{chen.gray:2002}.

In order to validate the assumptions made by the various theoretical
approaches, it is nevertheless important to test the accuracy of the
theoretical predictions. To this end, computer simulations are ideal,
because they, in principle, probe the phase behavior of the model system
without resorting to approximations. With inexpensive computer power
readily available nowadays, several groups have taken the opportunity to
investigate the IN transition by means of simulations
\cite{dijkstra.roij.ea:2001, bolhuis.frenkel:1997, bates:1997, bates:1998,
allen:2000, allen:2000*b, al-barwani.allen:2000, mcdonald:2000}. An
example of this approach is \olcite{bolhuis.frenkel:1997}, where the
coexistence properties of the bulk isotropic and nematic phases of hard
spherocylinders are carefully mapped out using Gibbs ensemble Monte Carlo
\cite{panagiotopoulos:1987}. These simulations generally recover the
Onsager limit for long rods, while for shorter rods pronounced deviations
show up \cite{bolhuis.frenkel:1997}. Unfortunately, the Gibbs ensemble
cannot be used to measure $\gin$, which is the aim of this work.

To obtain $\gin$ in simulations, different techniques must be used. One
such technique is based on the anisotropy of the pressure tensor. In
\olcite{mcdonald:2000}, this method is applied to suspensions of
ellipsoids with axial ratio $\kappa=A/B=15$, where $A$ is the length of
the symmetry axis of the ellipsoids, and $B$ that of the transverse axis.
The corresponding interfacial tension is $0.006 \pm 0.005 \; \kb T / B^2
\approx 0.09 \; \kb T / AB$ if a hard interaction potential is used, and
$0.011 \pm 0.004 \; \kb T / B^2 \approx 0.165 \; \kb T/ AB$ using a soft
potential. Note that the anisotropy of the pressure tensor is very small,
and therefore difficult to measure accurately in practice, as indicated by
the error bars.

In \olcite{akino.schmid.ea:2001}, again for (soft) ellipsoids with
$\kappa=15$, a value of the interfacial tension $\gin = 0.016 \pm 0.002 \;
\kb T / B^2 \approx 0.24 \; \kb T / AB$ is reported. This result was
obtained by measuring the capillary broadening of the IN interface.
According to capillary wave theory \cite{rowlinson.widom:1982}, the mean
squared amplitudes of the capillary fluctuations are proportional to
$1/\gin$, and this can be used to obtain the interfacial tension.
Unfortunately, capillary wave theory is only valid in the long wavelength
limit, such that very large system sizes are required. Moreover, if, as in
\olcite{akino.schmid.ea:2001}, periodic boundary conditions are used, two
interfaces will be present in the simulation box. Since $\gin$ is very
small, large capillary fluctuations can occur, and one needs to be aware
of interactions between the two interfaces.

Clearly, in order to obtain $\gin$ more accurately, much more computer
power or different simulation techniques are required. Recent advances in
grand canonical sampling methods \cite{yan.de-pablo:2000,
virnau.muller:2004} have enabled accurate measurements of the interfacial
tension in simple fluids \cite{potoff.panagiotopoulos:2000, gozdz:2003},
and complex fluids such as polymer solutions \cite{virnau.muller.ea:2004}
and colloid--polymer mixtures \cite{vink.horbach:2004*1}. The aim of this
paper is to apply these techniques to the IN transition in a system of
soft spherocylinders, and to extract the corresponding phase diagram and
the interfacial tension. Simulations in the grand canonical ensemble offer
a number of advantages over the more conventional methods discussed
previously. More precisely, in grand canonical simulations, both the
coexistence properties can be probed, as in the Gibbs ensemble, as well as
the interfacial properties. Additionally, finite--size scaling methods are
available which can be used to extrapolate simulation data to the
thermodynamic limit \cite{binder:1982, landau.binder:2000,
bruce.wilding:1992, kim.fisher.ea:2003}. It has been demonstrated that
grand canonical ensemble simulations combined with novel finite size
scaling algorithms can yield results of truly impressive accuracy
\cite{kim.fisher.ea:2003}.
	
This article is structured as follows: First, we introduce the soft
spherocylinder model used in this work. Next, we describe the grand
canonical Monte Carlo method, and explain how the coexistence properties,
and the interfacial tension are obtained. Finally, we present our results,
followed in the last section by a discussion and an outlook to future
work.

\begin{figure}
\begin{center}
\includegraphics[clip=,width=\figwidth]{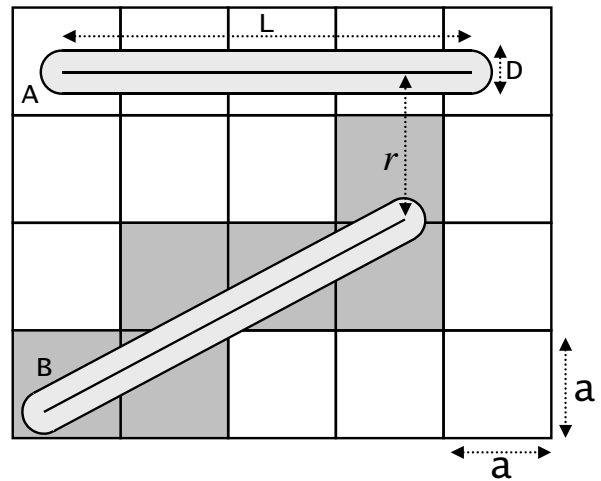}

\caption{\label{rods} Two--dimensional representation of the simulation
model of this work. The liquid crystals are modeled as soft
spherocylinders with elongation $L$ and diameter $D$. Two rods A and B
interact via the pair potential of \eq{eq:pot}, which is a function of
their minimum distance $r$ only. If the rods overlap, the system pays a
constant energy cost $\epsilon$. To speed up the determination of overlap,
the simulation box is subdivided into cubic cells with edge length $a$,
see details in text.}

\end{center}
\end{figure}

\section{Model}

In this study, the particles are modeled as repulsive soft spherocylinders of
elongation $L$ and diameter $D$. For numerical convenience, a very simple
potential has been chosen. The interaction between two rods $A$ and $B$, is
given by a pair potential of the form
\begin{eqnarray}
\label{eq:pot}
  V_{AB} (r) &=&
  \begin{cases}
  \epsilon & r < D, \\
  0 & {\rm otherwise},
  \end{cases}
\end{eqnarray}
with $r$ the distance between two line segments of length $L$, see
\fig{rods}. The total energy is thus proportional to the number of
overlaps in the system. In this work, the rod diameter $D$ is taken as
unit of length, and $\kb T$ as unit of energy. The strength of the
potential is set to $\epsilon = 2$. Note that in the limit $\epsilon \to
\infty$, this model approaches a system of infinitely hard rods.

To study the IN transition, we typically use the density and the rod alignment
as order parameters. Note that both the isotropic and the nematic phase are
fluid phases, in the sense that long--range positional order of the centers of
mass is absent. In the nematic phase, however, there is orientational order
where, on average, the rods point in one direction (called the director). In the
isotropic phase, on the other hand, there is no orientational order. Since the
density of the nematic phase is slightly higher than of the isotropic phase, we
may use the particle number density $\rho=N/V$ to distinguish between both
phases, with $N$ the number of rods in the system, and $V$ the volume of the
simulation box. Following convention, we also introduce the reduced density
$\rho^\star = \rho / \rhocp$, with $\rhocp = 2 / (\sqrt{2} + (L/D)\sqrt{3})$ the
density of regular close packing of hard spherocylinders. Orientational order is
as usual measured by the $S_2$ order parameter, defined as the maximum
eigenvalue of the orientational tensor $Q$:
\begin{equation}
\label{eq:s2}
 Q_{\alpha\beta} = \frac{1}{2 N} \sum_{i=1}^N
   \left( 3 u_{i\alpha} u_{i\beta} - \delta_{\alpha\beta} \right).
\end{equation}
Here, $u_{i\alpha}$ is the $\alpha$ component ($\alpha = x,y,z$) of the
orientation vector $\vec{u}_i$ of rod $i$ (normalized to unity), and
$\delta_{\alpha\beta}$ is the Kronecker delta. In case of orientational order,
such as in the nematic phase, $S_2$ assumes a value close to one, while in the
disordered isotropic phase, $S_2$ is close to zero.

\section{Simulation method}
\label{Simu}

The simulations are performed in the grand canonical ensemble. In this
ensemble, the volume $V$, the temperature $T$, and the chemical potential
$\mu$ of the rods are fixed, while the number of rods $N$ inside the
simulation box fluctuates. Insertion and removal of rods are attempted
with equal probability, and accepted with the standard grand canonical
Metropolis rules, given by $A(N \to N+1) = \min \left[1, \frac{V}{N+1} e^{
-\beta\Delta E + \beta \mu} \right]$ and $A(N \to N-1) = \min \left[1,
\frac{N}{V} e^{ -\beta\Delta E - \beta \mu} \right]$, with $\Delta E$ the
energy difference between initial and final state, and $\beta=1 / \kb T$
\cite{landau.binder:2000, frenkel.smit:2001}. The simulations are
performed in a three dimensional box of size $\lx \times \ly \times \lz$
using periodic boundary conditions in all directions. In this work, we fix
$\lx=\ly$, but we allow for elongation in the remaining direction $\lz
\geq \lx$. Moreover, to avoid double interactions between rods through the
periodic boundaries, we set $\lx > 2L$.

\begin{figure}
\begin{center}
\includegraphics[clip=,width=\figwidth]{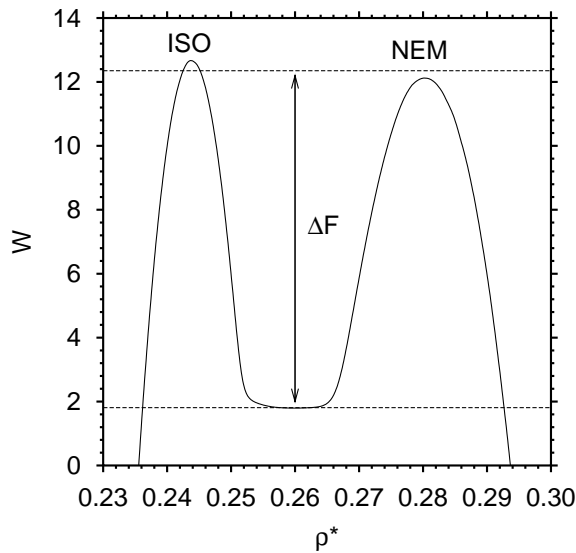}

\caption{\label{bimodal} Coexistence distribution $W = \kb T \ln \Pn$ of
the isotropic to nematic transition in a system of soft rods interacting
via \eq{eq:pot} with $\epsilon=2$ and $L/D=15$. The low density peak
corresponds to the isotropic phase (ISO), the high density peak to the
nematic phase (NEM), and the barrier $\Delta F$ to the free energy
difference between the two phases ($\Delta F$ is given by the average peak
height as measured from the minimum in between the peaks). The above
distribution was obtained using box dimensions $\lx=2.1L$ and $\lz=8.4L$.
The coexistence value of the chemical potential reads $\mu=5.15$ and was
obtained using the equal area criterion described in the text.}

\end{center}
\end{figure}

During the simulations, we measure the probability $\Pn$, defined as the
probability of observing a system containing $N$ rods. Note that the shape
of the distribution will depend on the rod elongation $L/D$, the
temperature $T$, and the chemical potential $\mu$. Moreover, there may be
finite--size effects, introducing additional dependences on the box
dimensions $\lx$ and $\lz$. At phase coexistence, the distribution $\Pn$
becomes bimodal, with two peaks of equal area, one located at small values
of $N$ corresponding to the isotropic phase, and one located at high
values of $N$ corresponding to the nematic phase. A typical coexistence
distribution is shown in \fig{bimodal}, where the logarithm of $P(N)$ is
plotted. Coexistence is determined using the equal area rule
\cite{muller.wilding:1995}. At coexistence, the equal area rule implies
that $\int_0^\avg{N} \Pn {\rm d} N = \int_\avg{N}^\infty \Pn {\rm d} N$,
with $\avg{N}$ the average of the full distribution $\avg{N} =
\int_0^\infty N \Pn {\rm d} N$, where we assume that $\Pn$ has been
normalized to unity $\int_0^\infty \Pn {\rm d} N = 1$. The coexistence
density of the isotropic phase follows trivially from the average of $\Pn$
in first peak $\riso = (2/V) \int_0^\avg{N} N \Pn {\rm d} N$, and
similarly for the nematic phase $\rnem = (2/V) \int_\avg{N}^\infty N \Pn
{\rm d} N$, where the factors of two are a consequence of the
normalization of $\Pn$.

\begin{figure}
\begin{center}
\includegraphics[clip=,width=\columnwidth]{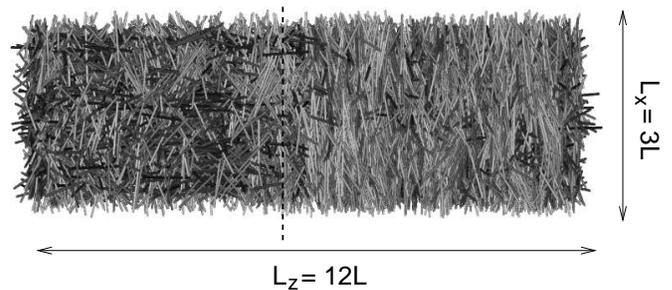}

\caption{\label{snapshot} Snapshot of a system of soft spherocylinders at IN
coexistence. The spherocylinders are shaded according to their orientation. On
the left side of the dashed line the system is isotropic, on the right side it
is nematic. The second interface coincides with the boundaries of the box in the
elongated direction.}

\end{center}
\end{figure}

The interfacial tension $\gin$ is extracted from the logarithm of the
probability distribution $W \equiv \kb T \ln \Pn$. Since $-W$ corresponds to the
free energy of the system, the average height $\Delta F$ of the peaks in $W$,
measured with respect to the minimum in between the peaks, equals the free
energy barrier separating the isotropic from the nematic phase. When the overall
density of the system is in the interval between the peaks $\riso \ll \rho \ll
\rnem$, coexistence between an isotropic and nematic domain is observed. A
snapshot of the system in this regime reveals a slab geometry, with one
isotropic region, and one nematic region, separated by an interface (because of
periodic boundary conditions, there are actually two interfaces). An example
snapshot is shown in \fig{snapshot}. Note that the director of the nematic phase
lies in the plane of the interfaces. This was the typical case for the snapshots
studied by us, and is consistent with the theoretical prediction that
in--plane alignment yields the lowest free energy. 

The barrier $\Delta F$ in \fig{bimodal} thus corresponds to the free energy cost
of having two interfaces in the system. Since, in this work, the box dimensions
are chosen such that $\lx=\ly$ and $\lz \geq \lx$, the interfaces will be
oriented perpendicular to the elongated direction, since this minimizes the
interfacial area, and hence the free energy of the system. The total interfacial
area in the system thus equals $2 \lx^2$. Since the interfacial tension is
simply the excess free energy per unit area, we may write
\begin{equation}
\label{eq:st}
  \gin(\lx) = \Delta F/ (2 \lx^2),
\end{equation}
with $\gin(L_x)$ the interfacial tension in a finite simulation box with lateral
dimension $L_x$ \cite{binder:1982}. To obtain the interfacial tension in the
thermodynamic limit, one can perform a finite size scaling analysis
\cite{binder:1982} to estimate $\lim_{\lx \to \infty} \gin(\lx)$. 
Alternatively, away from any critical point, the most dominant finite size
effects will likely stem from interactions between the two interfaces. In this
case, it is feasible to use an elongated simulation box with $\lz \gg \lx$, such
as in \fig{snapshot}. The advantage of using an elongated simulation box is that
interactions between the interfaces are suppressed. This enhances a flat region
in $W$ between the peaks, indicating that the interfaces are no longer
interacting, and that finite size effects will likely be small. In this work,
both approaches will be used.

If the free energy barrier $\Delta F$ is large, transitions between the
isotropic and the nematic phase become less likely, and the simulation will
spend most of the time in only one of the two phases. A crucial ingredient in
our simulation is therefore the use of a biased sampling technique. We use
successive umbrella sampling \cite{virnau.muller:2004} to enable accurate
sampling in regions where $\Pn$, due to the free energy barrier separating the
phases, is very small. Note also that phase coexistence is only observed if the
chemical potential $\mu$ is set equal to its coexistence value. This value is in
general not known at the start of the simulation, but it may easily be obtained
by using the equation $P(N|\mu_1) = P(N|\mu_0) e^{\beta (\mu_1 - \mu_0)N}$, with
$P(N|\mu_\alpha)$ the probability distribution $\Pn$ at chemical potential
$\mu_\alpha$. In the simulations, we typically set the chemical potential to
zero and use successive umbrella sampling to obtain the corresponding
probability distribution. We then use the above equation to obtain the desired
coexistence distribution, in which the area under both peaks is equal.

\section{Numerical Optimizations}

Most of the CPU time in our simulations is spend on calculating the distance $r$
between two line segments, see \fig{rods}. Naturally, one tries to minimize the
number of calls to the routine that determines the distance. To this end, we use
a cubic linked cell structure, which is schematically illustrated in \fig{rods}.
The crucial point is that the lattice constant $a$ is chosen such that $D < a <
L$. To determine if rod B in \fig{rods} overlaps with any of the other rods in
the system, it is sufficient to consider only those rods contained in the cubes
intersected by rod B (shaded gray), plus the rods contained in the nearest and
next--nearest neighbors of these cubes. Since the isotropic to nematic
transition occurs at low density, most cubes will be empty, resulting in a
substantial efficiency gain. Some CPU time is used for manipulating the linked
cell structure, but for large systems ($\approx N > 1500$) and long rods
($\approx L/D > 10$), the gain in efficiency is already a factor of five. Some
fine--tuning is required to obtain an optimal value of the lattice constant. We
found that $a \approx 0.2 L$ typically gives good results.

A further optimization concerns the calculation of the $S_2$ order parameter,
see \eq{eq:s2}. In a naive implementation, determining the orientational tensor
$Q$ involves an ${\cal O}(N)$ loop over all rods in the system. In our
implementation, the tensor elements of $Q$ are updated after each accepted Monte
Carlo move, which can be done at the cost of only a few additions and
multiplications. Since we keep the tensor elements updated throughout the
simulation, the ${\cal O}(N)$ loop of \eq{eq:s2} never needs to be carried out.
Finally, to determine the maximum eigenvalue of $Q$, we do not use a numerical
scheme, but instead use the exact expression for the roots of a third degree
polynomial. The advantage of this implementation is that the value of $S_2$ is
known exactly throughout the simulation, at a cost exceeding no more than one
percent of the total invested CPU time.

We conclude this section with a few benchmarks. For $\epsilon=2$ in
\eq{eq:pot}, we found that the acceptance rate of grand canonical
insertion is around 9 percent in the isotropic phase, and it decreases to
around 6 percent in the nematic phase. The acceptance rates are rather
insensitive to $L/D$. With the optimized implementation described in this
section, we can typically generate 5000--8000 accepted grand canonical
moves per second on a 2.2 GHz AMD Opteron processor.

\begin{figure}
\begin{center}
\includegraphics[clip=,width=\figwidth]{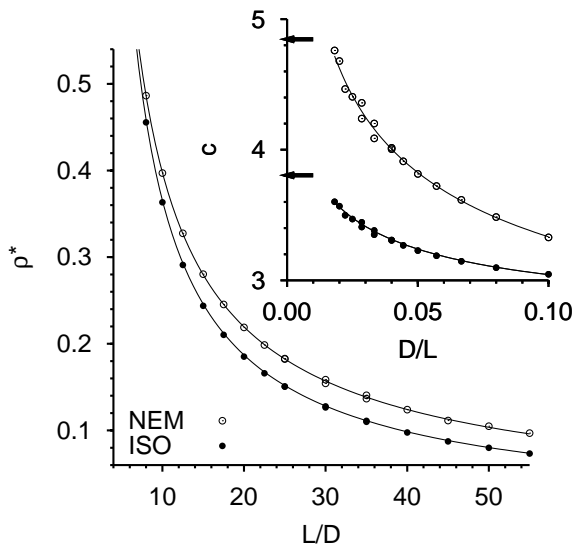}

\caption{\label{phase} Soft spherocylinder phase diagram of the IN
transition using $\epsilon=2$ in \eq{eq:pot}. Shown is the reduced
density $\rho^\star$ of the isotropic phase (closed circles) and of the
nematic phase (open circles) as function of $L/D$. The inset shows the
concentration variable $c$ as function of $D/L$ for both the isotropic and
the nematic phase. The lower and upper arrow in the inset mark the Onsager
limit $D/L \to 0$ for the isotropic and the nematic phase, respectively.
The lines connecting the points serve as a guide to the eye.}

\end{center}
\end{figure}

\section{Results}

\subsection{Phase diagram}

We first use our grand canonical Monte Carlo scheme to determine the IN
phase diagram of the soft spherocylinder system of \eq{eq:pot} using
$\epsilon = 2$. For several rod elongations $L/D$, we measured the
distribution $\Pn$, from which $\riso$ and $\rnem$ were obtained. The
system size used in these simulations is typically $\lx=\ly=2.1 L$ and
$\lz = 4.2 L$. In \fig{phase}, we plot the reduced density of the
isotropic and the nematic phase as function of $L/D$. We observe that the
phase diagram is qualitatively similar to that of hard spherocylinders
\cite{bolhuis.frenkel:1997}. The quantitative difference being that, for
soft rods, the IN transition is shifted towards higher density. The inset
of \fig{phase} shows the concentration variable $c=\pi D L^2 \rho/4$ as a
function of $D/L$. For hard spherocylinders, Onsager theory predicts that
$c_{\rm ISO}=3.29$ and $c_{\rm NEM}=4.19$ in the limit of infinite rod
length, or equivalently $D/L \to 0$. In case of the soft potential of
\eq{eq:pot}, these values must be multiplied by $(1-e^{-\beta
\epsilon})^{-1} \approx 1.16$ for $\epsilon = 2$. In the inset of
\fig{phase}, the corresponding limits are marked with arrows. As in
\olcite{bolhuis.frenkel:1997}, we observe that the simulation data for the
isotropic phase smoothly approach the Onsager limit, while the nematic
branch of the binodal seems to overshoot the Onsager limit. This we
attribute to equilibration problems. To simulate the IN transition in the
limit $D/L \to 0$, large system sizes are required, and it becomes
increasingly difficult to obtain accurate results. To quantify the
uncertainty in our measurements, additional independent simulations for
rod elongation $L/D=25$, $30$, and $35$ were performed. The corresponding
data are also shown in \fig{phase}. For $L/D \geq 30$, we observe
significant scatter, while for $L/D \leq 25$, the uncertainty is typically
smaller than the symbol size used in the plots.

\subsection{Interfacial tension}

Next, the interfacial tension $\gin$ of the IN interface is determined for
$L/D=10$ and $L/D=15$. Unfortunately, the system size used to compute the
phase diagram in the previous section, was insufficient to accurately
extract the interfacial tension because no flat region between the peaks
in $\Pn$ could be distinguished. This indicates that the interfaces are
still strongly interacting. To properly extract the interfacial tension,
much larger systems turned out to be required. In this case, care must be
taken in the sampling procedure. Many sampling schemes, especially the
ones that are easy to implement such as successive umbrella sampling, put
a bias on the density only. Such schemes tend to ``get stuck'' in
meta--stable droplet states when the system size becomes large
\cite{virnau.muller.ea:2004}. As a result, one may have difficulty
reaching the state with two parallel interfaces, in which case \eq{eq:st}
cannot be used.

\begin{figure}
\begin{center}
\includegraphics[clip=,width=\figwidth]{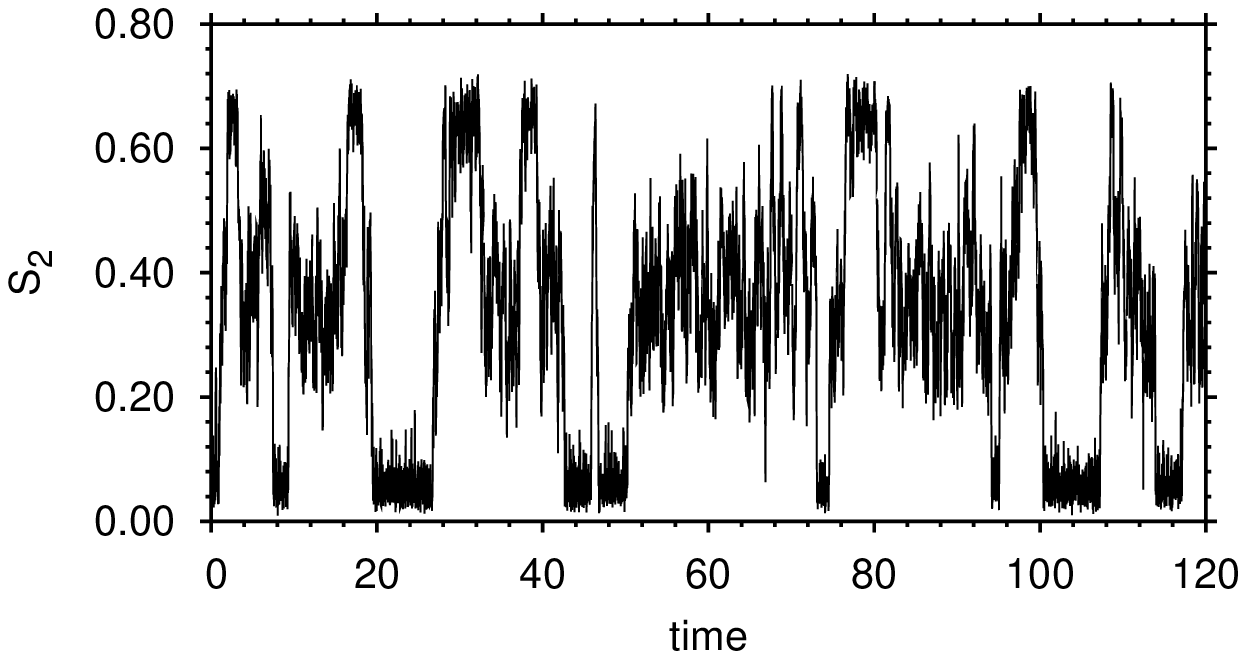}\\
\includegraphics[clip=,width=\figwidth]{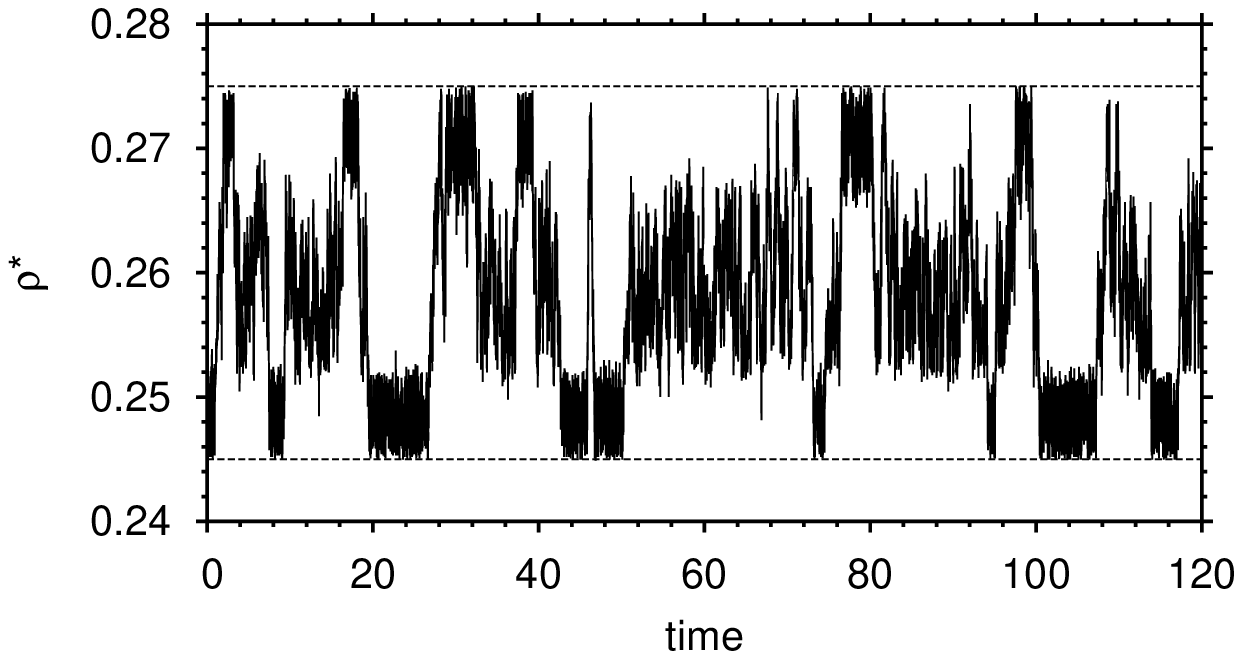}

\caption{\label{time} Monte Carlo time series of a biased grand canonical
simulation. The top frame shows the $S_2$ order parameter as a function of
the invested CPU time, the lower frame the reduced density, with CPU time
expressed in hours on a 2.6 GHz Pentium. During the simulation, the
reduced density was confined to the interval $0.245 < \rho^\star <0.275$,
as indicated by the horizontal lines in the lower figure. The data were
obtained using $L/D=15$, $\epsilon=2$, $\lx=2.1L$ and $\lz=8.4L$, which
are the same parameters as used in \fig{bimodal}.}

\end{center}
\end{figure}

Therefore, for large systems, one must carefully check the validity of the
simulation results. To this end, we occasionally inspect simulation
snapshots. For sufficiently elongated simulation boxes $\lz \gg \lx$ and
at densities inside the coexistence region $\riso \ll \rho \ll \rnem$, we
indeed observe two planar interfaces oriented perpendicular to the
elongated direction, in accord with \fig{snapshot}. To further check the
consistency of the measured distributions $\Pn$, we performed a number of
additional grand canonical simulations using a biased Hamiltonian of the
form ${\cal H} = {\cal H}_0 + W$, with ${\cal H}_0$ the Hamiltonian of the
real system defined by \eq{eq:pot} and $W=-\kb T \ln \Pn$. If the measured
$\Pn$ is indeed the equilibrium coexistence distribution of the real
system, a simulation using the biased Hamiltonian should visit the
isotropic and the nematic phase equally often on average
\cite{wilding:2001, virnau.muller:2004}. This is illustrated in the top
frame of \fig{time}, which shows the $S_2$ order parameter as a function
of the elapsed simulation time during one such biased simulation. Indeed,
we observe frequent transitions between the isotropic ($S_2 \sim 0$) and
the nematic phase ($S_2 \sim 1$). Also shown in \fig{time} is the
corresponding time series of the reduced density. In case a perfect
estimate for $\Pn$ could be provided, the measured distribution in the
biased simulation will become flat in the limit of long simulation time.
The deviation from a flat distribution can be used to estimate the error
in $\Pn$, or alternatively, to construct a better estimate for $\Pn$. The
latter approach was in fact adopted by us. First, successive umbrella
sampling is used to obtain an initial estimate for $\Pn$. This estimate is
then used as input for a number of biased simulations using the modified
Hamiltonian, and improved iteratively each time.

\begin{figure}
\begin{center}
\includegraphics[clip=,width=\figwidth]{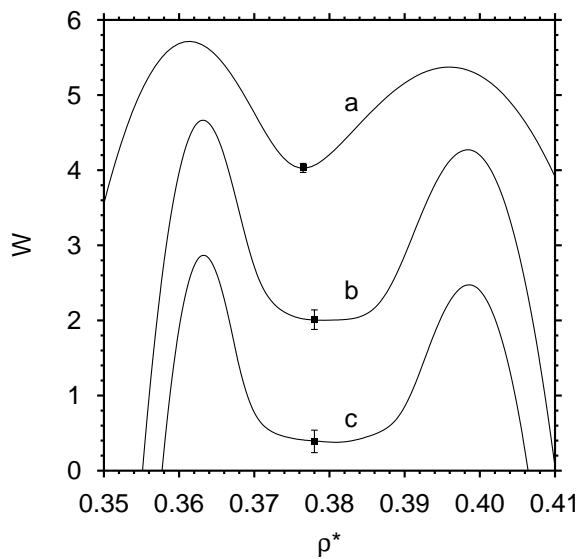}

\caption{\label{asp} Coexistence distributions $W = \kb T \ln \Pn$ of soft
spherocylinders with $L/D=10$ and $\epsilon=2$ for various system sizes.
In each of the above distributions, the lateral box dimension was fixed at
$\lx=\ly=2.3L$, while the perpendicular dimension was varied: (a)  
$\lz=2.3 L$; (b) $\lz=10.35 L$; (c) $\lz=13.8 L$. The corresponding free
energy barriers $\Delta F$ are: (a) $1.52 \pm 0.05$; (b) $2.47 \pm 0.13$;
(c) $2.29 \pm 0.15$, in units of $\kb T$. The error bars indicate the
magnitude of the scatter in $\Delta F$ for a number of independent
measurements.}

\end{center}
\end{figure}

To obtain the interfacial tension, the most straightforward approach is to
fix the lateral box dimensions at $\lx=\ly$, and to increase the elongated
dimension $\lz \gg \lx$ until a flat region between the peaks in the
distribution $\Pn$ appears. For soft spherocylinders of elongation
$L/D=10$, the results of this procedure are shown in \fig{asp}. Indeed, we
observe that the region between the peaks becomes flatter as the
elongation of the simulation box is increased. Unfortunately, even for the
largest system that we could handle, the region between the peaks still
displays some curvature. In other words, the interfaces are still
interacting, indicating that even more extreme box elongations are
required. Ignoring this effect, and applying \eq{eq:st} to the largest
system of \fig{asp}, we obtain for the interfacial tension $\gin = 0.0022
\; \kb T/D^2$. For rod elongation $L/D=15$, the distribution of the
largest system that we could handle is shown in \fig{bimodal}. The height
of the barrier reads $\Delta F=10.6 \; \kb T$, and the corresponding
interfacial tension $\gin = 0.0053 \; \kb T/D^2$.

\begin{figure}
\begin{center}
\includegraphics[clip=,width=\figwidth]{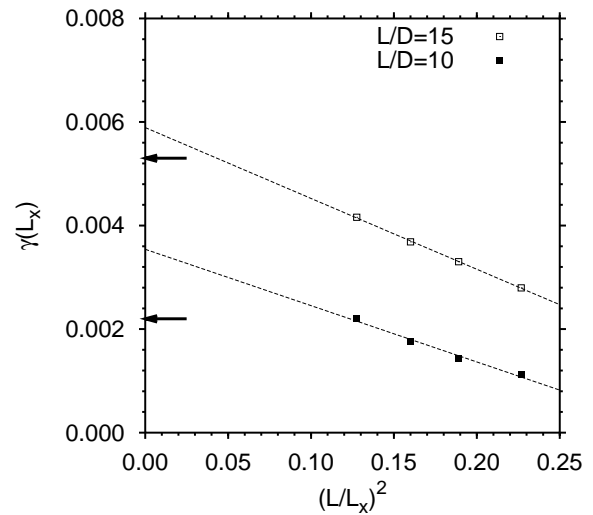}

\caption{\label{fss} Finite size extrapolation of the IN interfacial
tension of soft spherocylinders with $\epsilon=2$ and rod elongation
$L/D=10$ and 15. Shown is the interfacial tension of the finite system
$\gamma(\lx)$ in units of $\kb T / D^2$, measured in a cubic system with
edge $\lx$, as a function of $(L/\lx)^2$. Lines are linear fits to the
data using \eq{eq:fss} with $b=0$. The upper (lower) arrow indicates the
estimate of $\gin$ obtained using the method of \fig{asp} for
$L/D=15$~(10).}

\end{center}
\end{figure}

An alternative method to obtain the interfacial tension is to perform a
finite size scaling analysis. Following \olcite{binder:1982}, the
interfacial tension $\gamma(\lx)$ in a cubic system with edge $\lx$, shows
a systematic $\lx$ dependence that can be written as:
\begin{equation}
  \label{eq:fss}
  \gamma(\lx) = \gamma_\infty + a / \lx^2 + b \ln(\lx) / \lx^2,
\end{equation}
with $\gamma_\infty$ the interfacial tension in the thermodynamic limit
(assuming periodic boundary conditions and dimensionality $d=3$). In
general, the constants $a$ and $b$ are not known. However, recent
theoretical arguments \cite{kaupuzs:2004} suggest that in three
dimensions, the logarithmic term should vanish, implying $b=0$. To
estimate $\gamma_\infty$, we used \eq{eq:st} to measure $\gamma(\lx)$ for
a number of different system sizes. We then used \eq{eq:fss} to
extrapolate these measurements to the thermodynamic limit, assuming $b=0$.
For soft spherocylinders, the results of this procedure are summarized in
\fig{fss}. Shown is the interfacial tension of the finite system as a
function of $(L/\lx)^2$. The data seem reasonably well described by
\eq{eq:fss}, as is indicated by the fits. The corresponding estimates for
the interfacial tension are $\gin=0.0035 \; \kb T/D^2$ and $\gin=0.0059 \;
\kb T/D^2$, for $L/D=10$ and 15, respectively.

\begin{table*}

\caption{\label{tab:prop} Bulk properties of the coexisting isotropic and
nematic phase in a system of soft spherocylinders interacting via \eq{eq:pot}
with $\epsilon=2$ and rod elongation $L/D=10$ and 15. Listed are the reduced
density $\rho^\star$ and the normalized number density $\rho LD^2$ of the
isotropic and the nematic phase. Also listed is the interfacial tension $\gin$
of the IN interface, obtained using finite size scaling, expressed in various
units to facilitate the comparison to other work. The error bar in the 
latter quantity indicates the uncertainty of the fit in \fig{fss}.}

\begin{ruledtabular}
\begin{tabular}{c|cc|cc|ccc}
\multicolumn{1}{c|}{$L/D$} & 
\multicolumn{2}{c|}{isotropic phase} & 
\multicolumn{2}{c|}{nematic phase} & 
\multicolumn{3}{c}{interfacial tension $\gin$} \\
   & $\rho^\star$ & $\rho LD^2$ 
& $\rho^\star$ & $\rho LD^2$ 
& $\frac{\kb T}{D^2}$ & $\frac{\kb T}{LD}$ & $ \frac{\kb T}{(L+D)D}$  \\ \hline
10 & 0.363  & 0.388 & 0.397 & 0.424 & $0.0035 \pm 0.0003$ & 0.035 & 0.039  \\
15 & 0.244  & 0.267 & 0.280 & 0.307 & $0.0059 \pm 0.0001$ & 0.089 & 0.094
\end{tabular}
\end{ruledtabular}

\end{table*}

For comparison, the arrows in \fig{fss} mark the interfacial tension as
obtained using the previous method of \fig{asp}. Clearly, there is some
discrepancy. The problem related to the first method is that the system
size was not sufficient to completely suppress interface interactions.
Moreover, the lateral $\lx$ dimension was also rather small, so there may
still be finite size effects in this dimension. Hence, we believe the
finite size scaling results to be more reliable. The latter estimates are
listed in Table~\ref{tab:prop}, together with the coexisting phase
densities, which effectively summarizes the main results of this work. To
our knowledge, this is the first study to report a systematic finite size
scaling analysis of the IN interfacial tension. The results of \fig{fss}
seem reasonable, but simulations of larger systems are clearly needed, in
order to confirm the validity of \eq{eq:fss} in systems of elongated
particles. The advantage of the present simulation approach is that the
statistical errors are small, and that finite size effects are clearly
visible as a result.

\section{Discussion}

In this section, we compare our findings to other work. More precisely, we
consider (1) theoretical treatments within the Onsager approximation, (2)
theoretical treatments beyond the Onsager approximation, and (3) other
simulations. For reasons outlined in the introduction, we do not compare
to experimental data.

It is clear from the phase diagram of \fig{phase} that the Onsager limit is not
recovered until for very large rod elongation, exceeding at least $L/D=40$. As a
result, our estimates for the interfacial tension differ profoundly from Onsager
predictions. Typically, $\gin$ in our simulations is four times lower compared
to Onsager estimates. Note that our simulations also show that $\gin$ increases
with $L/D$, towards the Onsager result, so there seems to be qualitative
agreement. However, to properly access the Onsager regime, additional
simulations for large elongation $L/D$ are required. Unfortunately, as indicated
by the scatter in the data of \fig{phase}, and also in
\olcite{bolhuis.frenkel:1997}, such simulations are tremendously complicated. It
is questionable if present simulation techniques are sufficiently powerful to
extract $\gin$ with any meaningful accuracy in the Onsager regime.

If we compare to the theory of \olcite{velasco.mederos:2002}, which goes
beyond the Onsager approximation and should therefore be more accurate for
shorter rods, we observe better agreement. For $L/D=10$, the theory
predicts $\gin = 0.0877 \; \kb T / (L+D)D$, which still differs from our
result by a factor of approximately 2. For $L/D=15$, however, a naive
interpolation of the data in \olcite{velasco.mederos:2002} yields $\gin
\approx 0.1 \; \kb T / (L+D)D$, which exceeds our result by only 6\%. Note
that \olcite{velasco.mederos:2002} considers hard spherocylinders, whereas
our work is based on soft spherocylinders. The simulations of
\olcite{mcdonald:2000} on ellipsoids suggest that the interfacial tension
increases, when switching from a hard to a soft potential. The good
agreement we observe with \olcite{velasco.mederos:2002} should therefore
be treated with some care.

As mentioned in the introduction, computer simulations of soft ellipsoids
with $\kappa=15$ yield interfacial tensions of $\gin = 0.011 \pm 0.004 \;
\kb T/B^2$ and $\gin = 0.016 \pm 0.002 \; \kb T/B^2$ \cite{mcdonald:2000,
akino.schmid.ea:2001}. For $L/D=15$, our result for soft spherocylinders
is considerably lower. Obviously, spherocylinders are not ellipsoids, and
this may well be the source of the discrepancy. Note also that the shape
of the potential used by us is different from that of
Refs.~\onlinecite{mcdonald:2000} and \onlinecite{akino.schmid.ea:2001}.

In summary, we have performed grand canonical Mon-te Carlo simulations of
the IN transition in a system of soft spherocylinders. By measuring the
grand canonical order parameter distribution, the coexistence densities as
well as the interfacial tension were obtained. In agreement with
theoretical expectations and other simulations, ultra--low values for the
interfacial tension $\gin$ are found. Our results confirm that for short
rods, the interfacial tension, as well as the coexistence densities, are
considerably lower than the Onsager predictions. This demonstrates the
need for improved theory to describe the limit of shorter rods, which is
required if the connection to experiments is ever to be made. In the
future, we hope to extend our simulation method to the case of hard
spherocylinders. Note that grand canonical simulations of hard particles
are challenging, because the acceptance rate for insertion is typically
very low. We are currently investigating different biased sampling
techniques in order to improve efficiency. Also the investigation of the
structural properties of the IN interface is in progress.

\acknowledgments

We are grateful to the Deutsche Forschungsgemeinschaft (DFG) for support
(TR6/A5) and to K. Binder, M. M\"uller, P. van der Schoot, and R. van Roij
for stimulating discussions. We also thank G. T. Barkema for suggesting
some of the numerical optimizations used in this work. T.~S.~was supported
by the Emmy Noether program of the DFG.

\bibstyle{revtex}
\bibliography{mainz}

\end{document}